\documentclass[12pt,preprint]{aastex}

\shorttitle{TeV Survey of FSRQs}
\shortauthors{Falcone et al.}

\begin{document}


\title{A Search for TeV Gamma-Ray Emission from High-Peaked Flat Spectrum Radio Quasars Using the Whipple Air-Cherenkov Telescope}


\author{A.D. Falcone\altaffilmark{1,2}, 
I.H. Bond\altaffilmark{3}, 
P.J. Boyle\altaffilmark{4}, 
S.M. Bradbury\altaffilmark{3}, 
J.H. Buckley\altaffilmark{6}, 
D. Carter-Lewis\altaffilmark{7}, 
O. Celik\altaffilmark{8}, 
W. Cui\altaffilmark{1}, 
M. Daniel\altaffilmark{7}, 
M. D'Vali\altaffilmark{3}, 
I. de la Calle Perez\altaffilmark{3,20}, 
C. Duke\altaffilmark{9}, 
D.J. Fegan\altaffilmark{10}, 
S.J. Fegan\altaffilmark{11}, 
J.P. Finley\altaffilmark{1},
L.F. Fortson\altaffilmark{4,5}, 
J. Gaidos\altaffilmark{1}, 
S. Gammell\altaffilmark{10}, 
K. Gibbs\altaffilmark{11}, 
G.H. Gillanders\altaffilmark{12}, 
J. Grube\altaffilmark{3}, 
J. Hall\altaffilmark{13}, 
T.A. Hall\altaffilmark{14},
D. Hanna\altaffilmark{15}, 
A.M. Hillas\altaffilmark{3}, 
J. Holder\altaffilmark{3}, 
D. Horan\altaffilmark{11},
A. Jarvis\altaffilmark{8}, 
G.E. Kenny\altaffilmark{12}, 
M. Kertzman\altaffilmark{16},
D. Kieda\altaffilmark{13}, 
J. Kildea\altaffilmark{15}, 
J. Knapp\altaffilmark{3}, 
K. Kosack\altaffilmark{6},
H. Krawczynski\altaffilmark{6}, 
F. Krennrich\altaffilmark{7}, 
M.J. Lang\altaffilmark{12}, 
S. LeBohec\altaffilmark{7},
E. Linton\altaffilmark{4}, 
J. Lloyd-Evans\altaffilmark{3}, 
A. Milovanovic\altaffilmark{3},
P. Moriarty\altaffilmark{17}, 
D. Muller\altaffilmark{4}, 
T. Nagai\altaffilmark{13}, 
S. Nolan\altaffilmark{1},
R. Ong\altaffilmark{8}, 
R. Pallassini\altaffilmark{3}, 
D. Petry\altaffilmark{18},
F. Pizlo\altaffilmark{1}, 
B. Power-Mooney\altaffilmark{10},
J. Quinn\altaffilmark{10}, 
M. Quinn\altaffilmark{17}, 
K. Ragan\altaffilmark{15}, 
P. Rebillot\altaffilmark{6},
P.T. Reynolds\altaffilmark{19}, 
H.J. Rose\altaffilmark{3}, 
M. Schroedter\altaffilmark{11},
G. Sembroski\altaffilmark{1}, 
S.P. Swordy\altaffilmark{4}, 
A. Syson\altaffilmark{3},
K. Tyler\altaffilmark{1}, 
V.V. Vassiliev\altaffilmark{8},
S.P. Wakely\altaffilmark{4}, 
G. Walker\altaffilmark{13}, 
T.C. Weekes\altaffilmark{11}, 
J. Zweerink\altaffilmark{8}}


\altaffiltext{1}{Physics Department, Purdue University, West Lafayette, IN 47907}
\altaffiltext{2}{current address: Whipple Observatory, Amado, AZ 85645}
\altaffiltext{3}{Dept. of Physics, University of Leeds, Leeds LS2 9JT, Yorkshire, England, UK}
\altaffiltext{4}{University of Chicago, Chicago, IL 60637, USA}
\altaffiltext{5}{Astronomy Department, Adler Planetarium and Astronomy Museum, Chicago, IL 60637, USA}
\altaffiltext{6}{Dept. of Physics, Washington University, St. Louis, MO 63130, USA}
\altaffiltext{7}{Dept. of Physics and Astronomy, Iowa State University, Ames, IA 50011, USA}
\altaffiltext{8}{Dept. of Physics, University of California, Los Angeles, CA 90095, USA}
\altaffiltext{9}{Physics Dept., Grinnell College, Grinell, IA 50112, USA} 
\altaffiltext{10}{Physics Dept., National University of Ireland, Belfield, Dublin 4, Ireland} 
\altaffiltext{11}{Fred Lawrence Whipple Observatory, Harvard-Smithsonian CfA, Amado, AZ 85645, USA} 
\altaffiltext{12}{National University of Ireland, Galway, Ireland} 
\altaffiltext{13}{University of Utah, Salt Lake City, UT 84112 USA} 
\altaffiltext{14}{Dept. of Physics and Astronomy, University of Arkansas, Little Rock, AR 72204 USA} 
\altaffiltext{15}{Dept. of Physics, McGill University, Montreal, QC H3A 2T8 Canada} 
\altaffiltext{16}{DePauw University, Greencastle, IN 46135 USA} 
\altaffiltext{17}{Galway-Mayo Institute of Technology, Galway, Ireland} 
\altaffiltext{18}{NASA Goddard Space Flight Center, Code 661, Greenbelt, MD 20771 USA} 
\altaffiltext{19}{Dept. of Physics, Cork Institute of Technology, Cork, Ireland }
\altaffiltext{20}{Now at Dept. of Physics, Oxford University, UK}


\begin{abstract}

    Blazars have traditionally been separated into two broad categories based upon their optical emission characteristics.  Blazars with faint or no emission lines are referred to as BL Lacertae objects (BL Lacs), and blazars with prominent, broad emission lines are commonly referred to as flat spectrum radio quasars (FSRQs).  The spectral energy distribution of FSRQs has generally been thought of as being more akin to the low-peaked BL Lacs, which exhibit a peak in the infrared region of the spectrum, as opposed to high-peaked BL Lacs (HBLs), which exhibit a peak in UV/X-ray region of the spectrum.  All blazars that are currently confirmed as sources of TeV emission fall into the HBL category.  Recent surveys have found several FSRQs that exhibit spectral properties, particularly the synchrotron peak frequency, similar to HBLs.  These objects are potential sources of TeV emission according to several models of blazar jet emission and the evolution of blazars.  Measurements of TeV flux or flux upper limits could impact existing theories explaining the links between different blazar types and could have a significant impact on our understanding of the nature of objects that are capable of TeV emission.  In particular, the presence (or absence) of TeV emission from FSRQs could confirm (or cast doubt upon) recent evolutionary models that expect intermediate objects in a transitionary state between FSRQ and BL Lac.  The Whipple 10 meter imaging air-Cherenkov gamma-ray telescope is well suited for TeV gamma-ray observations.  Using the Whipple telescope, we have taken data on a small selection of nearby(z$<$0.1 in most cases), high-peaked FSRQs.  Although one of the objects, B2 0321+33, showed marginal evidence of flaring, no significant emission was detected.  The implications of this paucity of emission and the derived upper limits are discussed.

\end{abstract}


\section{Introduction}

Blazars, which exhibit the most extreme properties of all active galactic nuclei (AGNs), have traditionally been divided into two categories.  These categories are separated from one another based upon their optical line emission properties.  BL Lac objects have either no emission lines or weak and narrow emission lines, with the typical definition requiring rest frame equivalent widths less than 5 \AA \citep{sti91,per98}, while flat spectrum radio quasars (FSRQs) exhibit broad optical emission lines.  Both classes of objects have extreme natures, characterized by a wide range of flux and spectral variability on many timescales.  Objects from both classes also exhibit high levels of optical polarization.

The broadband radiation spectrum of blazars consists of two distinct components: a synchrotron spectrum component that spans radio to optical-ultraviolet wavelengths (and to X-rays for high-frequency peaked objects) and a high-energy component that can extend from the X-ray to the very high energy (VHE) gamma-ray (300 GeV $<$ E $<$ 100 TeV) regime.  Observationally, the spectra appear to have two distinct broad peaks when plotted in a $\nu$$f_{\nu}$ representation \citep{pad95,fos98,ghi98}.  The emission in the first peak of the spectral energy distribution (SED) is generally considered to be synchrotron emission from relativistic electrons.  The most widely invoked models which attempt to explain the higher energy emission of the second SED peak fall into the category of leptonic models.  These leptonic models posit that the X-ray to gamma-ray emission is produced by inverse Compton scattering of lower energy photons by beamed relativistic electrons.  The low energy photon fields could arise from synchrotron continuum photons within the jet (e.g. \citet{koe81}), or they could arise from ambient photons from the accretion disk which enter the jet directly (e.g. \citet{der92}) or after scattering or reprocessing (e.g. \citet{sik94}).  In addition to these leptonic models, hadronic models also attempt to explain the second component of the SED.  These models involve proton-initiated cascades (e.g. \citet{man93}) and/or proton synchrotron radiation \citep{mue01, aha00}, as well as synchrotron emission from secondary muons and pions\citep{mue03}.

In the blazar sequence proposed by \citet{fos97} and \citet{ghi98}, the first SED peak (the synchrotron peak) is at low energies for objects with a high bolometric luminosity, and the synchrotron peak is at high energies, sometimes in excess of 100 keV, for objects with low bolometric luminosity.  In this sequence, FSRQs have their peaks at the lowest frequencies, and BL Lac objects have a first peak at higher frequencies.  BL Lac objects are further divided into two categories: low-peaked BL Lacs (LBLs) exhibit a first peak at IR/optical frequencies, and high-peaked BL Lac objects (HBLs) exhibit a first peak at UV/X-ray frequencies.

With the advent of more recent deep X-ray and radio surveys, it is becoming clear that there exist objects that do not fall within the confines of this sequence.  Some FSRQ objects that exhibit intermediate to high frequency spectral peaks, like their BL Lac cousins, are being discovered.  \citet{pad97} created a catalog based on early surveys and found that $\approx$17$\%$ of the FSRQs in the sample had SEDs similar to HBLs, rather than LBLs.  \citet{per98} and \citet{lan01} have found that $\approx30\%$ of the FSRQs found in DXRBS have X-ray to radio luminosity ratios characteristic of HBLs, while still retaining broad and luminous emission lines characteristic of FSRQs.  The recent surveys attempt to bridge the gap between previous X-ray and radio selected objects, and they indicate the presence of a population of FSRQs with HBL-like SEDs, referred to as HFSRQs by \citet{per98}.  The possibility that these HFSRQs could produce GeV/TeV gamma-ray emission similar to their HBL cousins is explored in the observations described in this paper.

Although the catalog of confirmed and statistically significant ($>5\sigma$) VHE gamma-ray sources now includes 8 objects, 6 of which are AGN, the only type of AGN to be decisively detected at this time are HBLs \citep{wee03} (note: there are 18 VHE objects if one includes unconfirmed, yet published, sources).  FSRQs with high frequency peaks offer an opportunity to expand the VHE gamma-ray catalog and simultaneously constrain the nature of HFSRQs, as well as their associated emission mechanisms.  By determining the categories of objects that can (and can not) emit VHE gamma rays, the nature of the local medium required for such high energy acceleration can be characterized.

The remainder of this article is organized as follows.  Section 2 will identify and describe the chosen HFSRQ objects that are candidate GeV/TeV gamma-ray sources.  Section 3 will describe the Whipple telescope and the observations performed with it.  Section 4 will present the results of these observations, and Section 5 will discuss the implications of these results.

\section{Candidate Source Properties}

Candidate sources of TeV emission were chosen from lists of FSRQs based upon several criteria, including spectral characteristics at radio to X-ray wavelengths, the location of the object in the sky, and the redshift of the object.  The objects selected for observation are shown in Table 1.  These candidates were selected from the source lists published by \citet{per00} and \citet{pad02} following the application of several selection criteria.  Only objects with a high X-ray flux ($>$10$^{-12}$ erg cm$^{-2}$ s$^{-1}$ in the 0.1-2.4 keV energy band) were selected.  Objects were also selected by looking at their radio to X-ray spectral energy distribution to look for signs of a particularly high frequency of the first SED peak when plotted in a $\nu$f$_{\nu}$ representation.  The synchrotron peaks for the chosen candidates are in the range bounded by ${\sim}$few${\times}$10$^{14}$ Hz and ${\sim}2{\times}$10$^{16}$ Hz.  All of the sources chosen for this study have a high X-ray to radio luminosity ratio, as well as effective spectral indices that are more typical of HBLs than classical FSRQs.  Another important requirement that was placed on the candidate sources was that they be easily observable by the Whipple Observatory, which is restricted to observations during clear and moonless nights of Northern Hemisphere sources.

The last restriction that was generally placed on the source candidate list was that they be nearby (z$<$0.1).  Sources at higher redshifts begin to have a significant fraction of their GeV/TeV emission absorbed due to pair production from the interaction of the high energy photon and infrared photons from the diffuse extragalactic background \citep{gou67, vas00}.  At this time, the most distant confirmed source of TeV photons is at a redshift of z=0.129 (H1426+428), which places it at an optical depth of $\tau{\approx}1$ for 400 GeV photons \citep{dej02,hor02}.  All of the candidate sources chosen for this study, with the exception of RGB J1629+401, are at redshifts less than z=0.1.  According to \citet{pad02}, RGB J1629+401 has an exceptionally high first peak in its SED ($\sim2\times10^{16}$ Hz), so it was included in this survey, despite its redshift of z=0.272.

\begin{deluxetable}{cccccc}
%\tabletypesize{\scriptsize}
\tablecaption{Candidate FSRQ Source Properties \label{tbl-1}}
\tablewidth{0pt}
\tablehead{\colhead{Source} & \colhead{z} & \colhead{RA} & \colhead{Dec} & \colhead{F$_{X-ray}$\tablenotemark{a,b}} & \colhead{F$_{radio}$ (mJy)\tablenotemark{c}}}
\startdata
B2 0321+33 & 0.062 & 03 24 41.1 & +34 10 46  & 6.6\tablenotemark{a} & 368 \\
PG 2209+184 & 0.070 & 22 11 53.8 & +18 41 52 & 8.4\tablenotemark{a} & 116 \\
WGA J0838+2453 & 0.028 & 08 38 11.1 & +24 53 45 & 6.5\tablenotemark{a} & 32\tablenotemark{d} \\
RGB J1413+436 & 0.090 & 14 13 43.7 & +43 39 45 & 4.5\tablenotemark{b} & 50 \\
RGB J1629+4008 & 0.272 & 16 29 01.3 & +40 08 00 & 9.0\tablenotemark{b} & 20 \\
\enddata

%% Text for table notes should follow after the \enddata but before
%% the \end{deluxetable}. Make sure there is at least one \tablenotemark
%% in the table for each \tablenotetext.

\tablenotetext{a}{0.3-3.5 keV flux (10$^{-12}$ erg cm$^{-2}$ s$^{-1}$) corrected for galactic absorption taken from \citet{per00}}
\tablenotetext{b}{0.1-2.4 keV flux (10$^{-12}$ erg cm$^{-2}$ s$^{-1}$) uncorrected for galactic absorption taken from \citet{per00} and \citet{lau98}}
\tablenotetext{c}{4.85 GHz radio flux taken from \citet{gre91}}
\tablenotetext{c}{4.85 GHz radio flux taken from \citet{con95}}
%\tablecomments{Occasionally, authors wish to append a short
%paragraph of explanatory notes that pertain to the entire table, but
%which are different than the caption.  Such notes should be placed in
%a {\tt tablecomments} command like this.}

\end{deluxetable}

\section{Observations}
Observations were performed throughout the 2001/2002 and 2002/2003 seasons at the Whipple Observatory, using the Whipple 10-m imaging air Cherenkov telescope (IACT).

\subsection{The Whipple 10-m Telescope and Data Analysis}

The Whipple telescope consists of a 10 meter Davies-Cotton reflector \citep{dav57} and a 490 pixel camera \citep{fin01} comprised of photomultiplier tubes (this analysis uses only the inner 379 pixels, which have $0.12^{\circ}$ spacing).  When a VHE gamma ray or cosmic ray hits the top of the atmosphere of the Earth, the resulting cascade of particles produces Cherenkov light as it propagates down through the atmosphere.  This light is collected by the 10-m mirror, and the resulting image that is formed on the camera is used to characterize the shower.  The parameters used for this characterization are: alpha, length, width, distance, size, max1, and max2 \citep{hil98, pun91}.  These parameters are described in Table 2, and the geometry of the parameters is shown in Figure 1.  Due to different scattering angles during propagation of electromagnetic showers and hadronic showers, as well as the isotropic distribution of cosmic ray shower directions of origin, these characteristic parameters can be used to differentiate between gamma-ray and hadronic primary particles.  By placing cuts on these parameters, the background due to the more numerous cosmic rays can be reduced, and thus, the gamma-ray signal-to-noise ratio can be increased.  These cuts are optimized each season by observing the Crab Nebula, which is a steady emitter of VHE gamma rays \citep{hil98}, and maximizing the signal to noise ratio (significance).  This procedure, of course, leads to a situation in which the analysis is optimized for a Crab-like spectrum.  For sources that are significantly softer or harder, one has to be resigned to a loss in sensitivity, or one must re-optimize the cuts for a different spectral shape.  The Crab is also used to calibrate the response of the instrument.  The peak response of the telescope during the 2001/2002 and 2002/2003 seasons was at $\approx$400 GeV for a Crab-like spectrum.

\begin{deluxetable}{ll}
\tablecaption{Image Parameters Used for Characterization of Showers \label{tbl-2}}
\tablewidth{0pt}
\tablehead{
\colhead{Parameter} & \colhead{Description} 
}
\startdata
length &  length (major axis) of shower image \\
width &  width (minor axis) of shower image \\
distance &  distance from shower image centroid to camera center \\
size &  sum of signals from all pixels in shower image \\
max1 &  largest signal recorded in any image pixel \\
max2 &  second largest signal recorded in any image pixel \\
alpha &  angle of shower major axis relative to line from camera center to image centroid \\

\enddata
\end{deluxetable}

\begin{figure}
\plotone{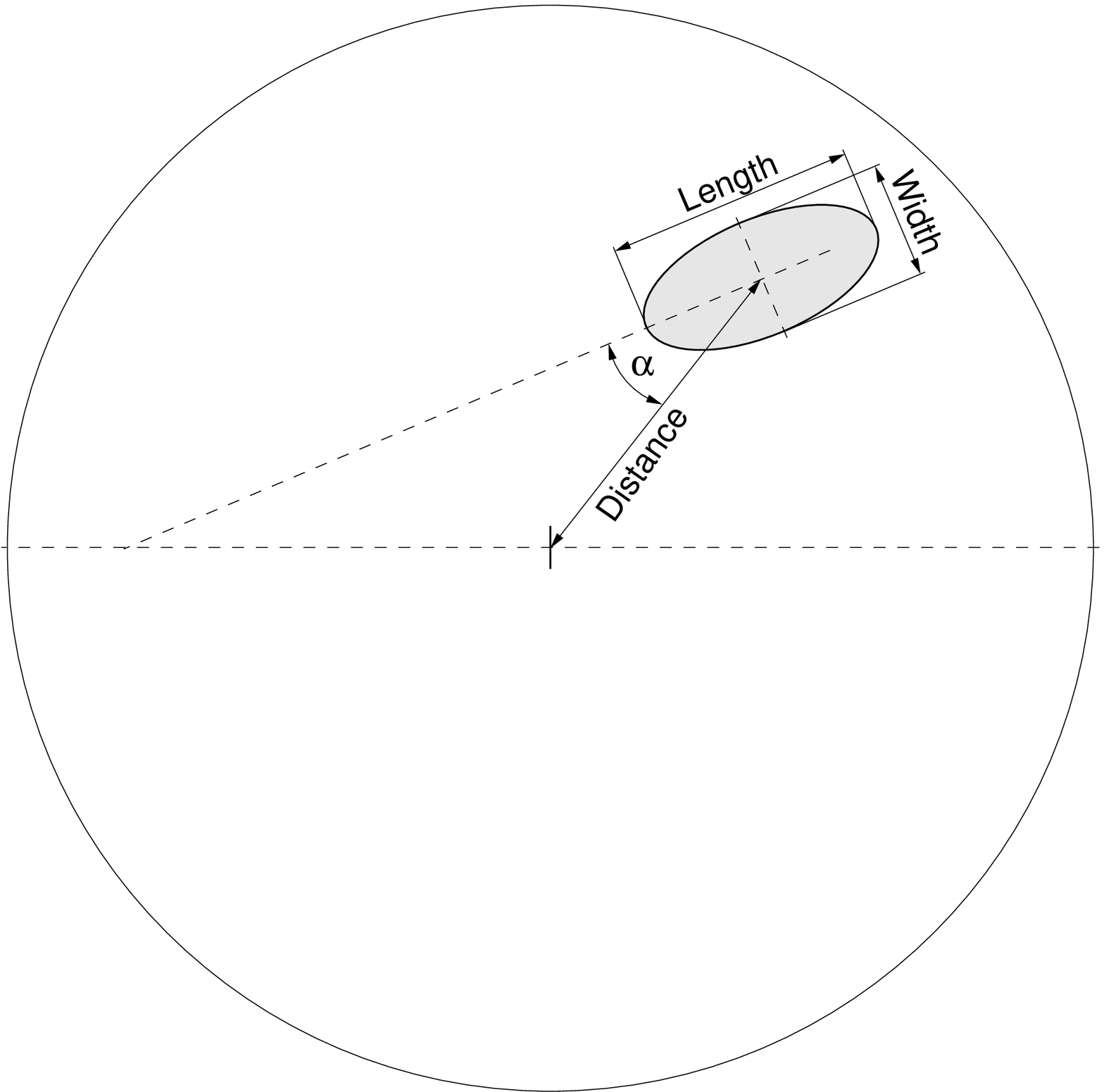}
\caption{Image characterization of air showers imaged at the focal plane of the telescope.  This cartoon shower projection has an elliptical shape that is typical of gamma ray and cosmic ray induced showers, and it has a large alpha angle, which implies that the field of view is not centered on the source.  \label{fig1}}
\end{figure}

The Whipple telescope is operated in two standard modes of operation.  The first of these, referred to as the ON/OFF mode, requires that the 28 minute run in which the telescope is pointed at the candidate source (ON) be followed by another 28 minute run (OFF) in which the telescope is offset in right ascension by 30 minutes such that it images a region of sky through the same range of elevation and azimuth as the original source pointing.  Since this OFF run is taken in nearly the same conditions as the ON run, without the candidate source in the field of view, it provides an independent measurement of the background.  When operating in the other mode, referred to as TRACKING, one observes only at the sky coordinates of the source.  This mode utilizes the large data set of OFF source observations taken throughout the season to estimate the background expected at small alpha angles (where the source counts are expected for point-like sources), as a function of the background counts at large alpha angles (where no source counts are expected).  This procedure takes advantage of the fact that gamma-ray events from the source will preferentially produce an image with a small alpha angle, while the more numerous background cosmic rays will be randomly oriented in the camera field of view.  While this method allows more on-source data to be taken, it requires an accurate calculation of the tracking ratio, which is the ratio of events with alpha angle between 20$^\circ$ and 65$^\circ$ to events with alpha angle between 0$^\circ$ and 15$^\circ$ in the absence of a gamma-ray source.  Since the analysis results are highly dependent on this tracking ratio, it must be calculated independently each season using OFF region data that have been taken when the telescope is operating in a state that is as close as possible to the state of operation when the TRACKING data is taken.  This ratio has been calculated and applied to this analysis.  For the 2001-2002 data, the tracking ratio is $0.312\pm0.003$; and for the 2002-2003 data, the tracking ratio is $0.3067\pm0.0035$.

Once the tracking ratio has been calculated using OFF source observations, the background for the ON source observations in the 0$^\circ$ to 15$^\circ$ alpha region, where all of the gamma-ray signal is expected, can be calculated.  This is done by using the tracking ratio to scale the background events in the 20$^\circ$ to 65$^\circ$ alpha region to obtain the expected background counts in the 0$^\circ$ to 15$^\circ$ alpha region.  After applying a simple propagation of error formula to this procedure, one obtains the following for the significance.

\begin{equation}
\sigma = \frac{N_{0^\circ-15^\circ}-r(N_{20^\circ-65^\circ})}{\sqrt{(N_{0^\circ-15^\circ})+r^2 N_{20^\circ-65^\circ}+({\Delta}r)^2 (N_{20^\circ-65^\circ})^2}},
\end{equation}
where $r\pm\Delta{r}$ is the tracking ratio and its associated statistical error.  An important characteristic of this method for computing significance is that it incorporates the error on the tracking ratio, which is a quantity that is not accommodated by the method of \citet{lim83}.  It was found by \citet{lim83} that this type of significance calculation tends to be more conservative than their reported method of calculating significance.  It should be noted that the error used to calculate this significance value is the statistical error; it does not include possible systematic errors.

A more detailed description of the standard analysis and the use of various cuts has been published previously in several articles including \citet{moh98}, \citet{rey93}, and \citet{cat98}.

\subsection{Source Observations}

Each of the candidate FSRQs was observed by the Whipple telescope, as shown in Table 3.  Two of the candidates were observed during the 2001/2002 season (B2 0321+33 and PG 2209+184), and three additional candidates were observed during the 2002/2003 season (WGA J0838+2453, RGB J1413+436, and RGB J1629+4008).  One source, B2 0321+33, was observed during both seasons since there was a marginal hint of flaring during the first season that merited a follow-up set of observations.  During both seasons, the image parameter cuts were optimized independently, but the optimized cuts were identical for the two seasons, as expected for a stable camera configuration.  Tracking ratios were calculated from a large number of OFF source runs taken during the relevant time frames.  All of the observations reported here were taken at zenith angles less than 35$^{\circ}$.

\begin{deluxetable}{ccccccc}
\tabletypesize\scriptsize
\tablecaption{Whipple Observations of Candidate FSRQ Sources \label{tbl-3}}
\tablewidth{0pt}
\tablehead{
\colhead{Source} & \colhead{ON$_{2001-2002}$}   & \colhead{ON$_{2002-2003}$}   &
\colhead{$\sigma$\tablenotemark{b}} & \colhead{upper limit\tablenotemark{c}} &
\colhead{upper limit\tablenotemark{d}} & \colhead{upper limit\tablenotemark{e}}\\
\colhead{ } & \colhead{(min)\tablenotemark{a}} & \colhead{(min)\tablenotemark{a}} & \colhead{ } & \colhead{(Crab)} & \colhead{(10$^{-11}$ erg cm$^{-2}$ s$^{-1}$)} & \colhead{(10$^{-11}$ erg cm$^{-2}$ s$^{-1}$)}
}
\startdata
B2 0321+33 & 320.3 & 331.5 & 1.56  & 0.10 & 0.52 & 0.68 \\
PG 2209+184 & 217.6 & 0 & 0.15 & 0.13 & 0.71 & 0.97 \\
WGA J0838+2453 & 0 & 781.6  & 0.15  & 0.05 & 0.29 & 0.32 \\
RGB J1413+436 & 0 & 535.4 & 0.75 & 0.06 & 0.35 & 0.53 \\
RGB J1629+4008 & 0 & 415.1 & 1.29 & 0.09 & 0.47 & 2.9 \\
\enddata

%% Text for table notes should follow after the \enddata but before
%% the \end{deluxetable}. Make sure there is at least one \tablenotemark
%% in the table for each \tablenotetext.

\tablenotetext{a}{ON refers to minutes spent on source during stable/clear weather conditions}
\tablenotetext{b}{significance calculated as described in text}
\tablenotetext{c}{95\% confidence level upper limits calculated using method of \citet{hel83}, expressed as a fraction of the Crab flux ($R_{Crab}=2.84\pm0.16$ in 2001-2002 season and $R_{Crab}=2.55\pm0.13$ in 2002-2003 season)}
\tablenotetext{d}{95\% confidence level upper limits for emission above $\approx$400 GeV, assuming a Crab-like spectrum as in \citet{hil98}; calculated using method of \citet{hel83}}
\tablenotetext{e}{95\% confidence level upper limits as described above, including the effect of absorption from the extragalactic background, using the baseline model of \citet{dej02}}

%\tablecomments{Occasionally, authors wish to append a short
%paragraph of explanatory notes that pertain to the entire table, but
%which are different than the caption.  Such notes should be placed in
%a {\tt tablecomments} command like this.}

\end{deluxetable}

\section{Results}

Significance values calculated using the method described above are shown in Table 3.  None of the sources exhibited significant steady emission of gamma rays over the integrated time frame of the observations.  Also shown in Table 3 are the $95\%$ upper limit values for steady emission, calculated using the method of \citet{hel83}.  The fifth and sixth columns of Table 3 are the upper limits for flux detected at Earth, and the last column shows the upper limits for the flux at the source.  This last column has been calculated using the optical depth of the extragalactic background for 400 GeV photons, as given by the baseline model of \citet{dej02}.

Although none of the FSRQs showed evidence for steady emission during the extended observations, one of them did exhibit marginal evidence of flaring during the 2001/2002 observing season.  On 22 Oct. 2001 (MJD 52204) two 28 minute observation runs were taken on B2 0321+33.  The average gamma-ray rate of these two runs was 0.46 $\pm$ 0.14 times that of the Crab, which corresponds to a significance of 3.3$\sigma$.  The peak rate was 0.62 $\pm$ 0.19 Crabs.  The light curve of data taken during that season is shown in Figure 2.  If the trials factor is derived using the 16 observation nights of FSRQs during this season (combined for all sources), then the post-trials significance of this rate increase is 2.5$\sigma$.  While this rate increase was not significant enough to claim a detection of a variable source, it was enough to merit the further observations that were taken of this candidate in the 2002/2003 season.

\begin{figure}
\plotone{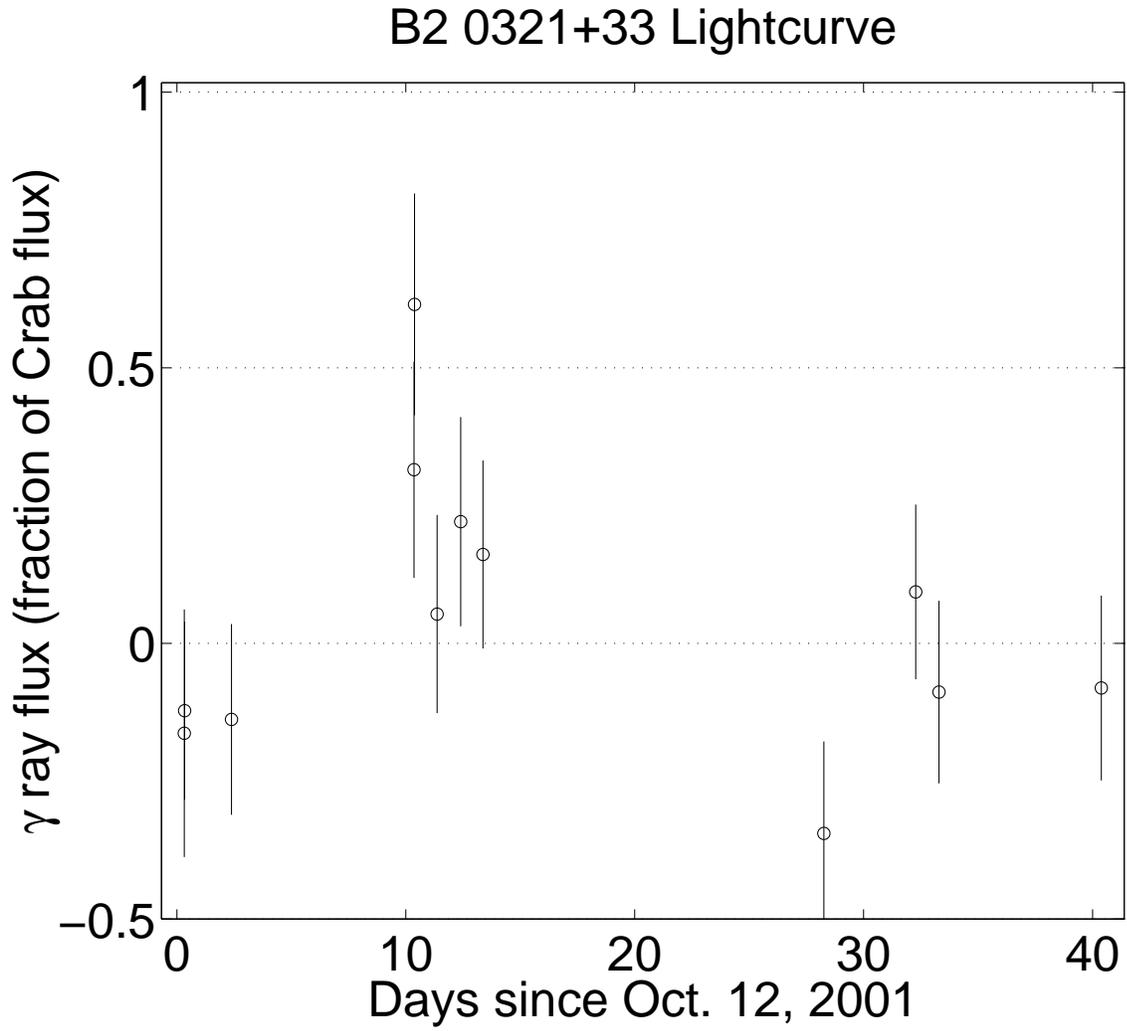}
\caption{Light curve of B2 0321+33 using all data taken in the 2001/2002 observing season.  Error bars are $1\sigma$ statistical errors.  \label{fig2}}
\end{figure}

\section{Discussion and Conclusions}

No significant emission has been detected from any of the candidate sources in this initial survey.  There was marginal evidence of a rate increase observed in the B2 0321+33 light curve, but the statistical significance of this increase is 2.5$\sigma$ (post-trials significance), which could be accounted for by a statistical fluctuation.  None of the other objects showed significant steady emission or any interesting features in their light curves.

The upper limits presented here are the first VHE gamma-ray upper limits published for HFSRQs; however, they do not severely constrain the emission models due to the large amount of uncertainty in the models (and, of course, due to limited sensitivity of the telescope).  Potential variability of the sources also leads to uncertainty in the expected flux.  Typically, FSRQs and BL Lacs are highly variable sources.  Any VHE gamma-ray flux detected from these objects would most likely be from the sum of many flares or from a large flare that happened to be caught during the time frame of the observation.  Applicable models of FSRQs have been made by \citet{pad02}.  In this paper, several high-peaked and intermediate HFSRQs were modelled using a one-zone synchrotron inverse Compton model that was developed by \citet{ghi02}.  This model includes synchrotron self-Compton, as well as seed photon contributions from the disk and the broad line region.  The inclusion of the latter components may be important for these sources, which are known to have broad line emission.  While \citet{per00} predicted that B2 0321+33, PG 2209+184, and RGB 1413+436 would be good TeV emission candidates, no modelling of the TeV flux was reported.

As a rough estimate of the expected flux, we have compared the SED of B2 0321+33 to those modelled by \citet{pad02}, and we find that it is similar to WGA J0546.6-6415, which has an estimated intrinsic very high energy gamma-ray flux of $\sim2\times10^{-12}$ erg cm$^{-2}$ s$^{-1}$ above 400 GeV.  This is below the value of our reported upper limit.  With the exception of RGB J1629+4008, the very high energy flux of the other objects is not as easy to predict within the context of published models.  More modelling is necessary.

\citet{pad02} directly modelled the SED of RGB J1629+4008, and they derived a synchrotron peak at $1.7\times10^{16}$ Hz and a very high energy gamma-ray flux of $\sim2\times10^{-14}$ erg cm$^{-2}$ s$^{-1}$ above 400 GeV.  No mention was made of absorption of gamma rays by the diffuse IR background.  This will produce a significant modification since the source is at a large redshift of z=0.272.  Using the model of \citet{dej02}, this redshift corresponds to an optical depth of $\approx$1.8 for 400 GeV gamma rays.  The intrinsic source flux upper limit derived here is $0.29\times10^{-12}$ erg cm$^{-2}$ s$^{-1}$ above 400 GeV, which is certainly higher than the predicted source flux so no strong constraints can be placed on the steady emission models for this object.  The particularly high synchrotron peak and the potential for increased flux during flaring prompted these observations, but no variability was detected during the observations.

\citet{cav02} and \citet{bot02} posit that the blazar sequence (i.e. the relationship of blazar type and synchrotron peak location to the bolometric luminosity) is fundamentally dependent on the accretion rate, \.{m}.  In this scenario, objects with large values of \.{m} are FSRQs and have large L$_{bol}$ (more than $\sim$10$^{46}$ erg s$^{-1}$); while objects with small values of \.{m} are BL Lacs and have small L$_{bol}$ ($\sim$10$^{44}$ erg s$^{-1}$).  Both of these models also predict an evolutionary scenario that progresses from FSRQs to BL Lacs, which leads to intermediate FSRQ objects that can have optical lines and large L$_{bol}$, along with a synchrotron peak more like that of BL Lacs.  These objects should be excellent candidates for VHE gamma-ray emission.  The fact that the observations presented here yield little or no VHE emission does not support this scenario, but these observations certainly do not rule out the proposed models for two reasons.  It is possible that these intermediate objects have optical depths that are large enough to prevent the escape of VHE emission.  It is also possible that the handful of objects in this initial survey were not in a high flaring state at the time of observation.  The inherent variability of blazars makes the detection of both BL Lac and FSRQ VHE emission difficult, as evidenced by the paucity of detected sources relative to the number of observed BL Lac candidates in past campaigns \citep{del03,hor04}.  More observations of these and other objects will be necessary.

Although no detection of VHE gamma-ray emission has been reported from this initial survey, dim or variable emission from these candidates, or emission from other similar candidates, is not ruled out.  This work should be continued with more sensitive instruments, such as the next generation of IACTs (VERITAS, HESS, MAGIC, and CANGAROO-III) that are beginning to come on line at the time of this writing \citep{kre04,hin04,lor04,kub04}.  More modelling of these objects is required in order to predict the high energy flux more accurately, and therefore, the potential of a given HFSRQ to be a gamma-ray source.  In addition, the continued progress of deeper X-ray to gamma-ray surveys (Swift, EXIST, INTEGRAL, GLAST) and X-ray imaging (Chandra, XMM) may reveal additional TeV FSRQ candidates with higher frequency peaks in the SED and/or increased high energy flux potential.

\acknowledgments
We acknowledge the technical assistance of E. Roache and J. Melnick.  This research is supported by grants from the U.S. Department of Energy, Enterprise Ireland, and PPARC in the UK.

%% The following command ends your manuscript. LaTeX will ignore any text


\begin{thebibliography}{}

\bibitem[Aharonian (2000)]{aha00} Aharonian, F. A. 2000, New Astronomy, 5, 377

\bibitem[B\"{o}ttcher \& Dermer (2002)] {bot02} B\"{o}ttcher, M. \& Dermer, C.D. 2002, \apj, 564, 86

\bibitem[Catanese et al. (1998)]{cat98} Catanese, M., et al. 1998, \apj, 501, 616

\bibitem[Cavaliere \& D'Elia (2002)] {cav02} Cavaliere, A. \& D'Elia, V. 2002, \apj, 571, 226 

\bibitem[Condon, Anderson, \& Broderick (1995)]{con95} Condon, J.J., Anderson, E., Broderick, J.J. 1995, \aj, 109, 2318C

\bibitem[Davies \& Cotton (1957)]{dav57} Davies, J.M., \& Cotton, E.S. 1957, Journal of Solar Energy, 1, 16

\bibitem[de Jager \& Stecker (2002)]{dej02} de Jager, O.C. \& Stecker, F.W. 2002, \apj, 566, 738

\bibitem[de la Calle P\'{e}rez et al. (2003)]{del03} de la Calle P\'{e}rez, I. et al. 2003, \apj, 599, 909

\bibitem[Dermer, Schlickeiser, \& Mastichiadis (1992)]{der92} Dermer, C.D., Schlickeiser, R., \& Mastichiadis, A. 1992, Astron. \& Astrophys., 256, L27

\bibitem[Finley et al. (2001)]{fin01} Finley, J.P. et al. 2001, in Proc. of 27th Inter. Cosmic Ray Conf. (Hamburg), OG 2.3, 199 

\bibitem[Fossati et al. (1998)]{fos98} Fossati, G. et al. 1998, Monthly Notices of the Royal Astronomical Society, 299, 433

\bibitem[Fossati et al. (1997)]{fos97} Fossati, G., Celotti, A., Ghisellini, G., \& Maraschi, L. 1997, Monthly Notices of the Royal Astronomical Society, 289, 136

\bibitem[Ghisellini, Celotti, \& Costamante (2002)]{ghi02} Ghisellini, G., Celotti, A., \& Costamante, L. 2002, Astron. \& Astrophys., 386, 833

\bibitem[Ghisellini et al. (1998)]{ghi98} Ghisellini, G. et al. 1998, Monthly Notices of the Royal Astronomical Society, 301, 451 

\bibitem[Gould \& Schreder (1967)]{gou67} Gould, R.P. \& Schreder, G.P. 1967, Phys. Rev., 155, 1408

\bibitem[Gregory \& Condon (1991)]{gre91} Gregory, P.C. \& Condon, J.J. 1991, \apjs, 75, 1011

\bibitem[Helene (1983)]{hel83} Helene, O. 1983, NIM, 212, 319

\bibitem[Hillas (1998)]{hil98} Hillas, A.M., et al. 1998, \apj, 503, 744

\bibitem[Hinton et al. (2004)]{hin04} Hinton, J., et al. 2004, New Astronomy Reviews, 48, 331

\bibitem[Horan et al. (2004)]{hor04} Horan, D., et al. 2004, \apj, in press

\bibitem[Horan et al. (2002)]{hor02} Horan, D., et al. 2002, \apj, 571, 753

\bibitem[Koenigl (1981)]{koe81} Koenigl, A. 1981, \apj, 243, 700

\bibitem[Krennrich et al. (2004)]{kre04} Krennrich, F., et al. 2004, New Astronomy Reviews, 48, 345

\bibitem[Kubo et al. (2004)]{kub04} Kubo, H., et al. 2004, New Astronomy Reviews, 48, 323

\bibitem[Landt et al. (2001)]{lan01} Landt, H., Padovani, P., Perlman, E.S., Giommi, P., Bignall, H., \& Tzioumis, A. 2001, MNRAS, 323, 757

\bibitem[Laurent-Muehleisen et al. (1998)]{lau98} Laurent-Muehleisen, S.A., Kollgaard, R.I., Ciardullo, R., Feigelson, E.D., Brinkmann, W., \& Siebert, J. 1998, \apjs, 118, 127

\bibitem[Li \& Ma (1983)]{lim83} Li, T. \& Ma, Y. 1983, \apj, 272, 317

\bibitem[Lorenz et al. (2004)]{lor04} Lorenz, E., et al. 2004, New Astronomy Reviews, 48, 339
 
\bibitem[Mannheim (1993)]{man93} Mannheim, K. 1993, Astron. \& Astrophys., 269, 67 

\bibitem[Mohanty et al. (1998)]{moh98} Mohanty, G., et al. 1998, Astroparticle Physics, 9, 15

\bibitem[Muecke et al. (2003)]{mue03} Muecke, A., Protheroe, R.J., Engel, R., Rachen, J.P., \& Stanev, T. 2003, Astroparticle Physics, 18, 593

\bibitem[Muecke and Protheroe (2001)]{mue01} Muecke, A. \& Protheroe, R. J. 2001, Astroparticle Physics, 15, 121

\bibitem[Padovani et al. (2002)]{pad02} Padovani, P., Costamante, L., Ghisellini, G., Giommi, P., \& Perlman, E. 2002, \apj, 581, 895

\bibitem[Padovani, Giommi, \& Fiore (1997)]{pad97} Padovani, P., Giommi, P., \& Fiore, F., Mem. Soc. Astron. Italiana, 68, 147

\bibitem[Padovani \& Giommi (1995)]{pad95} Padovani, P. \& Giommi, P., \apj, 444, 567

\bibitem[Perlman (2000)]{per00} Perlman, E.S. 2000, Proc. of GeV-TeV Gamma Ray Astrophysics Workshop, eds. Dingus, B.L., et al., 53

\bibitem[Perlman et al. (1998)]{per98} Perlman, E.S., Padovani, P., Giommi, P., Sambruna, R., Jones, L.R., Tzioumis, A., \& Reynolds, J. 1998, \apj, 115, 1253

\bibitem[Punch et al. (1991)]{pun91} Punch, M., et al. 1991, Proc. of 22nd Int. Cosmic Ray Conf.(Dublin), 1, 464

\bibitem[Reynolds et al. (1993)]{rey93} Reynolds, P., et al. 1993, \apj, 404, 206

\bibitem[Sikora, Begelman, \& Rees (1994)]{sik94} Sikora, M., Begelman, M.C., \& Rees, M.J. 1994, \apj, 421, 153  

\bibitem[Stickel et al. (1991)]{sti91} Stickel, M., Padovani, P., Urry, C.M., Fried, J.W., \& Kh\"ur, H. 1991, \apj, 374, 431

\bibitem[Vassiliev (2000)]{vas00} Vassiliev, V.V. 2000, Astroparticle Physics, 12, 217

\bibitem[Weekes (2003)]{wee03} Weekes, T.C. 2003, Proc. of 28th Int. Cosmic Ray Conf.(Tsukuba), {\it astro-ph/0312179}

\end{thebibliography}
\end{document}